\documentclass{PoS}

\title{Effects of the restoration of U$_A$(1) symmetry on pseudoscalar meson observables}

\ShortTitle{Effects of the restoration of U$_A$(1) symmetry on pseudoscalar meson observables}

\author{\speaker{Maria Ruivo}\\

        Departamento de F\'{i}sica, Universidade de Coimbra, Portugal\\

        E-mail: \email{maria@teor.fis.uc.pt}}

\author{P. Costa\\

        Departamento de F\'{i}sica, Universidade de Coimbra, Portugal\\

        E-mail: \email{pcosta@teor.fis.uc.pt}}

\author{C. A. de Sousa\\

        Departamento de F\'{i}sica, Universidade de Coimbra, Portugal\\

        E-mail: \email{celia@teor.fis.uc.pt}}

\author{Yu.L. Kalinovsky\\

        Laboratory of Information Technologies, Joint Institute for Nuclear Research, Dubna, Russia\\

        E-mail: \email{kalinov@nusun.jinr.ru}}

\abstract{
We investigate the restoration of chiral SU$(3)\otimes$SU$(3)$  and axial U$_A$(1) symmetries, at finite temperature and density, in the framework  of the three flavor Nambu-Jona-Lasinio model with anomaly. We implement a temperature (density) dependence of the anomaly coefficient motivated by  lattice results for the topological susceptibility and we discuss the  restoration of symmetries by analyzing the behavior of the mesonic chiral partners  and of  the mixing angles. 
The results indicate  that the axial part of the symmetry   is restored before the possible restoration  of the full U(3)$\otimes$U(3) chiral symmetry  can occur. 
}

\FullConference{International Europhysics Conference on High Energy Physics\\

                 July 21st - 27th 2005\\

                 Lisboa, Portugal}

\begin{document}

A challenging question in physics of strong interactions  has been the understanding of the physics of the low-lying hadron spectrum from the viewpoint of QCD dynamics and symmetries, the explicit and spontaneous breaking of chiral and axial symmetries playing an important role in this context.

Ultra-relativistic heavy-ion experiments are   expected to provide the strong interaction conditions which will lead to new physics. Restoration of symmetries and deconfinement are expected to occur, allowing for the search  of signatures of quark gluon plasma.  An interesting question  is whether  both  chiral SU$(N_f)\otimes$SU$(N_f)$  and axial U$_A$(1) symmetries are restored and which observables could exhibit manifestations  of  the possible restorations. The spectra of the low-lying scalar and pseudoscalar mesons should carry relevant information in this concern, since it is expected that the chiral partners get degenerate. Concerning specifically the U$_A$(1) symmetry, a sign of its restoration should be the vanishing of the topological susceptibility, which, in pure color SU(3) theory, can be linked to the $\eta'$ mass through the
Witten-Veneziano formula \cite{Veneziano}. In addition, since   the axial
anomaly causes flavor mixing, with the consequent violation of the
Okubo-Zweig-Iizuka (OZI) rule, the restoration  of axial symmetry should lead to the recovering of ideal mixing.

Several studies have been done linking the decrease with temperature of the topological susceptibility, $\chi$, with the restoration of the U$_A$(1) symmetry. Lattice results for $\chi$ exhibit a sharp drop of this quantity around the critical temperature \cite{lattice} and preliminary results at finite chemical potential also indicate a decrease of $\chi$ \cite{Alles}. In model calculations  the restoration of the U$_A$(1)  symmetry may be achieved by assuming that the anomaly coefficient is a dropping function of temperature, the form of this function being  inspired on  lattice results for  $\chi$ \cite{Ohta,Bielich} or based on phenomenological arguments \cite{alkofer}.

The present work aims at investigating the restoration of chiral and axial symmetries with temperature and density. We perform our calculations in the framework of an  extended  SU(3) Nambu--Jona-Lasinio model Lagrangian density that includes the 't Hooft determinant which breaks the U$_A(1)$ symmetry and we assume that the coefficient of this term is a dropping function of temperature (density).
The model Lagrangian is:
\begin{eqnarray}
{\mathcal L\,}&=& \bar q\,(\,i\, {\gamma}.\partial-\,\hat m)\,q
+ \frac{g_S}{2}\, \sum_{a=0}^8[\,{(\,\bar q\,\lambda^a\, q\,)}
^2+{(\,\bar q \,i\,\gamma_5\,\lambda^a\, q\,)}^2\,]  \nonumber \\
&+& g_D\,\{\mbox{det}\,[\bar q\,(1+\gamma_5)\,q] +\mbox{det}
\,[\bar q\,(1-\gamma_5)\,q]\}. 
\end{eqnarray}
By using a standard hadronization procedure, an effective action is obtained, leading to gap equations for the constituent quark masses and to meson propagators from which several observables are calculated \cite{costa}.
In the present work we follow the  methodology of \cite{Ohta,Bielich} and extract the temperature dependence of the anomaly coefficient $g_D$ from the topological susceptibility $\chi$, which is modeled as a Fermi function from lattice results \cite{lattice} and we extrapolate a similar approach for neutral cold quark matter in weak equilibrium (see Fig. 1, upper panels).
We discuss the possible restoration of  chiral SU$(3)\otimes$SU$(3)$ and U$_A$(1)  symmetries by analyzing the convergence of scalar and pseudoscalar mesons (chiral partners) and the  behavior of the   scalar and pseudoscalar mixing angles $(\theta_S\,\,,\theta_P)$

At temperatures around $T\approx200$ MeV the mass of the light quarks drops to the current quark mass, indicating a washed-out crossover. The strange quark mass, although decreasing, is still 2 times the strange current quark  mass at $T = 400$ MeV, meaning that chiral symmetry shows a slow tendency to get restored in the $s$ sector.
In fact, since $m_u=m_d<m_s$, the (sub)group SU(2)$\otimes$SU(2) is a much 
better symmetry of the NJL Lagrangian.
So, the effective restoration of the above symmetry  implies the degeneracy between the  chiral partners $(\pi^0,\sigma)$ and $(a_0,\eta)$ which, as it can be seen in Fig. 1- left panel, occurs around $T\approx250$ MeV. At $T\approx350$ MeV both $a_0$ and $\sigma$  mesons become degenerate with the $\pi^0$ and $\eta$ mesons, and the scalar and pseudoscalar mixing angles achieve the ideal values. This indicates  an effective restoration of both chiral and axial symmetries.  However, the $\eta^\prime$ and $f_0$ masses do not show a clear tendency to converge \cite{costaUA1}, which is a manifestation of the absence of restoration of chiral symmetry in the strange sector in the range of temperatures studied. 

The results obtained for finite density  are qualitatively similar, although some specific differences should be noticed. We analyze here the case of neutral quark matter in $\beta$--equilibrium, that undergoes a first order phase transition \cite{costa}.

Concerning the mixing angles for scalar and pseudoscalar mesons, $\theta_S$ and $\theta_P$, we observe that $\theta_S$, like in the finite temperature case,
starts at $16^{\circ}$ and increases up to the ideal mixing angle $35.264^{\circ}$. A different behavior is found for the angle $\theta_P$ that starts at $-5.8^{\circ}$ and, instead of getting the negative ideal value $-57.74^{\circ}$, as in the $T\neq 0$ case, changes sign at $\rho_B\approx4.8\rho_0$ and goes to the ideal mixing angle $35.264^{\circ}$, leading to a change of identity between $\eta$ and $\eta'$. 

The meson masses, as function of the density, are plotted in Fig. 1, right panel. The SU(2) chiral partners ($\pi^0,\sigma$) are  bound states and become 
degenerated at $\rho_B=3\rho_0$.
Concerning the SU(2) chiral partners ($\eta,a_0$), while the $a_0$ meson is always a purely non strange quark system, the $\eta$ has a strange component at $\rho_B = 0$ and, as the density increases, becomes less strange and degenerates with $a_0$ at $4.0\rho_0\leq\rho_B\leq4.8\rho_0$. In this range of densities ($\eta,a_0$) and ($\pi^0,\sigma$) are all degenerated. Suddenly the $\eta$ mass separates from the others becoming a purely strange state (see Fig. 1, right panel). This is due to the behavior of  $\theta_P$ that changes the sign and goes to $35.264^{\circ}$ at $\rho_B\approx4.8\rho_0$. On the other hand, the $\eta'$, that starts as an unbounded state and becomes bounded at $\rho_B>3.0\rho_0$, turns into a purely light quark system and degenerates with $\pi^0$, $\sigma$ and $a_0$ mesons at $\rho_B\approx4.8\rho_0$.
We conclude that the U$_A$(1) symmetry is effectively restored at $\rho_B>4\rho_0$ \cite{costaUA1,costaUA1big}.
In fact, the U$_A$(1) violating quantities show a tendency to vanish, which means that the four meson masses are degenerated and the topological susceptibility goes to zero.  This type of matter allows also the study of manifestations of restoration of symmetries on the charged mesons. We observe the convergence of charged pions and kaons with its scalar chiral partners, but differently from pions, kaons are not meaningfully affected by the restoration of  the U$_A$(1) symmetry \cite{costaUA1big}.

In summary, we have  implemented  a criterion which combines a lattice-inspired behavior of the topological susceptibility with the convergence of appropriate chiral partners to explore effective restoration of symmetries. We conclude that there is an effective restoration of chiral SU$(2)\otimes$SU$(2)$ and U$_A$(1)  symmetries, but  we do not observe the full restoration of the U(3)$\otimes$U(3) symmetry, although its axial part is restored at a moderate temperature (density).

\begin{figure}[t]
\includegraphics[width=7.75cm,height=11cm]{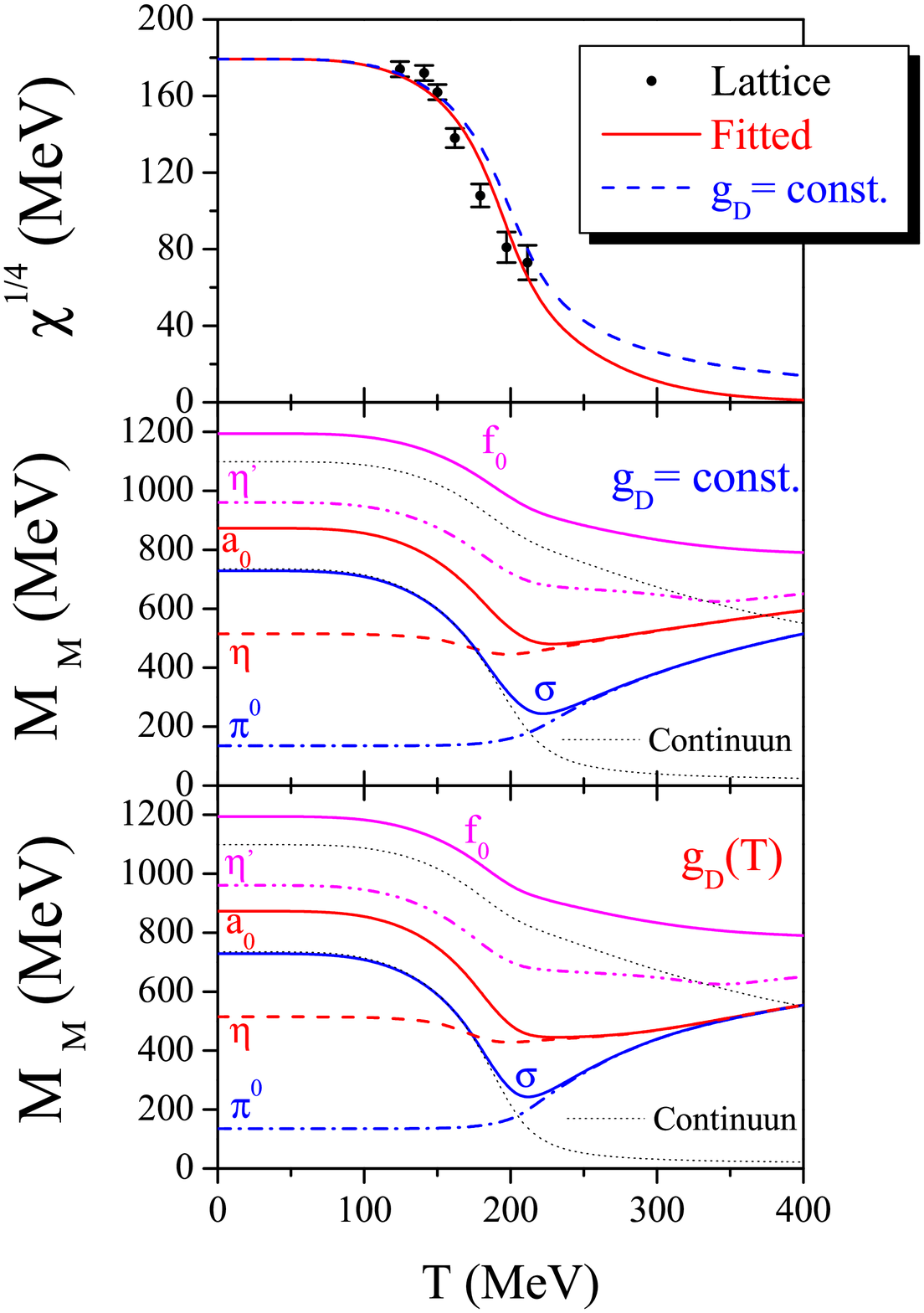}
\includegraphics[width=7.25cm,height=11cm]{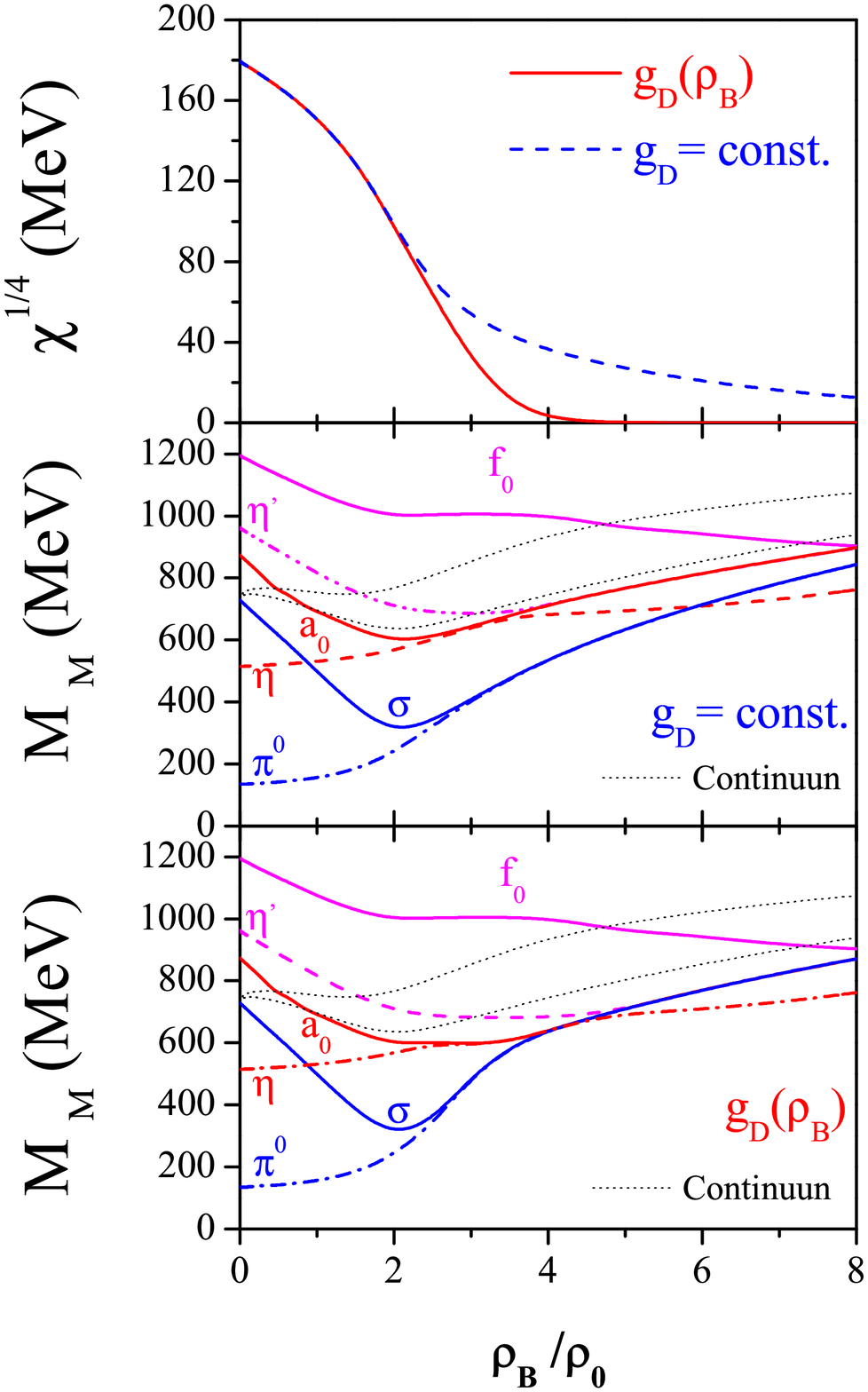}
\caption{Results at $T\neq 0$ (left panels) and at $\rho_B\neq0$  (right panels): topological susceptibility (upper panels) and neutral scalar and pseudoscalar meson masses with $g_D$ constant (middle panels) and with $g_D$ temperature (density) dependent (lower panels).}
\label{fig:dens}
\end{figure}

\vspace{1cm}
Work supported by grant SFRH/BD/3296/2000, FCT (P. Costa), CFT and by FEDER/FCT under project POCTI/FIS/451/94. 


\end{document}